\documentclass[twocolumn]{revtex4}
\usepackage{amssymb,epsfig,amscd}
\usepackage{graphicx}
\usepackage{dcolumn}
\usepackage{bm}
  
\usepackage{epsfig}

\newcommand{\dirac}{\partial\llap{$\diagup$\kern-2pt}}

\def\be{\begin{equation}} 
\def\ee{\end{equation}}
\def\bq{\begin{eqnarray}} 
\def\eq{\end{eqnarray}}

\begin{document}

\title{Formation of quark phases in protoneutron stars: the transition from the 2SC to the normal quark phase}
\author{G.~Pagliara}

\affiliation{Institut f\"{u}r Theoretische Physik, Ruprecht-Karls-Universit\"at,
   Philosophenweg 16, D-69120, Heidelberg, Germany}

\begin{abstract} 
We study the process of formation of quark phases in protoneutron
stars. After calculating the phase transition between nucleonic
matter and the 2SC phase at fixed entropy and lepton fraction, we show
that an unpairing transition between the 2SC phase and the normal
quark phase occurs for low lepton fractions. We then calculate the
process of diffusion of neutrinos in protoneutron stars and show that
for intermediate values of the mass of the star, the deleptonization
triggers the phase transition between the two quark phases after a
temporal delay of a few seconds. In less massive stars instead only
the normal quark phase is formed at the end of the deleptonization
stage. We also discuss the possible astrophysical implications of our
scenario.
\end{abstract}

\maketitle

\section{Introduction}
In the last years, a new very exciting theoretical discovery in QCD
concerns the low temperature and high density region of the phase diagram of strongly
interacting matter where quarks form a state of color
superconductivity. In
particular, the so called Color-Flavor-Locking (CFL) phase in which
all quarks are paired, was shown to be the ground state of strongly
interacting matter at asymptotically high densities
\cite{Alford:2007xm}. In turn, this opens the possibility that this
new quark phase appears in the core of neutron stars and affects their
properties.
A number of studies has been done to investigate the
structure of the QCD phase diagram for the conditions that are
realized in neutron stars: the requirements of charge neutrality and
equilibrium with respect to weak interactions, which must be fulfilled
in compact star matter, and the finite value of the mass of the strange quark 
split the values of the chemical potentials of the
different quark flavors thus allowing for the existence of many color
superconducting phases in the phase diagram as the 2SC phase, the
normal (unpaired) phase, the gapless phases and the crystalline phase
\cite{Steiner:2002gx,Ruester:2005jc,Blaschke:2005uj,Casalbuoni:2005zp,Warringa:2006dk}.
An important question concerns when and how these quark phases are
eventually formed in a neutron star. In
\cite{Steiner:2002gx,Ruester:2005ib,Sandin:2007zr,Lugones:2010gj} it was argued that
while the 2SC phase or the normal quark phase might be formed already
during the protoneutron star stage i.e. when matter is hot and lepton
rich, the CFL phase could appear only later when the star has almost
completed its deleptonization through the emission of neutrinos. The
possibility of forming different quark phases during the protoneutron
stars evolution could have very intriguing observable consequences as
proposed in \cite{Drago:2004vu,Drago:2005rc,Pagliara:2007ph,Drago:2008tb}
in connection with gamma-ray-bursts. 

To date, there are no detailed numerical simulations of the process of
formation of color superconducting phases in protoneutron stars: such
a calculation is indeed very complicated because, besides the equation
of state, it involves also the computation of the neutrino mean free
path in the different quark phases and it requires to solve
numerically the transport equations describing the diffusion of
neutrinos within the protoneutron star matter. A first preliminary
study was presented in \cite{Carter:2000xf}, where, besides the
calculation of the neutrino mean free path in color superconducting
quark matter, a simple modeling of the neutrino cooling was presented
in order to discuss the possible observable signatures of the
formation of a color superconducting phase in the star. The scenario
consists of a second order phase transition from the unpaired phase to
a superconducting phase: during the cooling, the temperature decreases
and at some point reaches the critical temperature and the peak of the
specific heat associated with the phase transition is met. The authors
argue that the phase transition would lead to a slower cooling of the
star for temperatures close to the critical temperature with possible
effects on the neutrino luminosity. This possibility could be indeed
realized in protoneutron stars if the color superconducting gap is of
the order of a few tens of MeV which lead to critical temperatures
comparable with the temperatures of protoneutron stars.

Here we will investigate a different scenario: at birth a protoneutron
star has a lepton fraction of the order of $Y_L = 0.4$ with a
corresponding electron/proton fraction of $Y_e = Y_p \sim 0.3$ at the center of the star. The
subsequent deleptonization causes a decrease of lepton and also
electron fractions: consequently also $Y_p$ decreases and matter
becomes more and more isospin asymmetric. If quark matter is present
in the core of the star at birth, it is likely to be in the 2SC phase
because matter is not very isospin asymmetric i.e.  the number of up
and down quarks are quite similar and the 2SC pairing is favored.  As
the deleptonization proceeds, the gradual increase of the isospin
asymmetry causes a stress on the Cooper pairs and a first order phase
transition from the 2SC to the unpaired quark phase is triggered at
some critical value of the asymmetry
\cite{Bedaque:1999nu,Kiriyama:2001ud,Lawley:2005ru,Ruester:2005ib,Pagliara:2010ii}.
Remarkably, the phase transition between the superconducting state and
the normal state in asymmetric fermionic systems has attracted much
attention in the last years also in connection with cold atoms
experiments (see for instance
\cite{Bedaque:2003hi,2006Sci...311..503P}).  By solving the neutrino
diffusion equation coupled with a quasi-static temporal evolution of
the structure of the star, we will show here that such a phase
transition between quark phases could indeed occur in protoneutron
star matter and possibly gives some observational effects in the
neutrino signal.

The paper is organized as follows: in Sec. II we calculate the
Equation of State (EoS) for protoneutron star matter by including the
pure quark and nucleonic phases and the mixed phase. In Sec. III we
present and discuss the results of our numerical simulation of the
neutrino diffusion and finally in Sec. IV we draw our conclusions.

\section{Equation of state of protoneutron star matter}
Matter in protoneutron stars is hot and lepton rich. At birth, the
initial conditions correspond, to good approximation, to uniform values
of the entropy per baryon $S/N=1$ and the lepton fraction $Y_L=0.4$
\cite{Prakash:1995uw,Steiner:2000bi}. Since the deleptonization reduces
the value of $Y_L$, we will calculate the EoS for $Y_L$ varying within the
range $0.04-0.4$. As we will explain in the next section, we disregard
here the effect of reheating of the star during deleptonization and we
will therefore keep $S/N$ fixed during the simulation.
\begin{figure}
    \begin{centering}
\epsfig{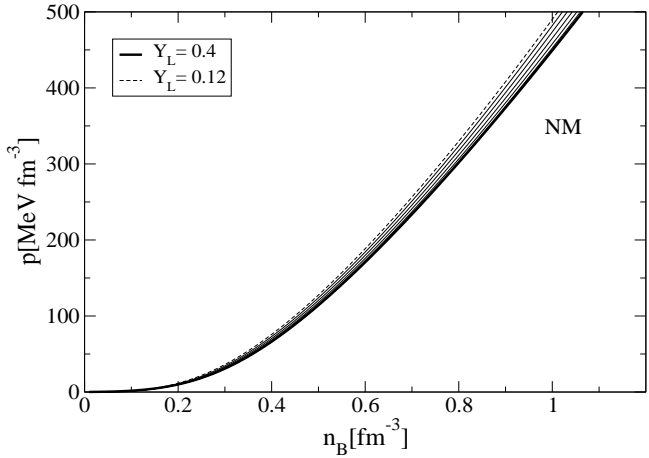}
    \caption{Pressure as a function of the baryon density for nucleonic matter for 
values of $Y_L$ ranging from $0.4$ (thick line) to $0.12$ (dashed line) with intervals
of $0.04$ (solid lines). At lower lepton fractions, which implies also lower electron and 
proton fractions, the nucleonic equation of state is stiffer because of the increasing contribution of the symmetry energy to the EoS.
\label{fig:pressnucl} }
   \end{centering}
\end{figure} 
To describe the phase transition from Nucleonic Matter (NM) to Normal
Quark matter (NQ) or 2SC matter, we use, as customary, two models: for
the low density nucleonic phase we adopt the relativistic mean field
model with the parameterization TM1 \cite{Shen:1998gq} and for the
high density quark phase a modified bag model which includes also 
color superconductivity in the 2SC phase used in
\cite{Pagliara:2010ii}. We limit our discussion here to two flavor
quark matter, with massless up and down quarks and we fix the model
parameters as follows: the intermediate value of the diquark coupling
is considered $G_D=3/4 G_S$ where $G_S$ is the scalar coupling and the
bag constant $B$ is fixed to $155$ MeV$^{1/4}$. We also include, in an effective
manner, the corrections to the quark pressure due to perturbative strong interactions 
\cite{Fraga:2001id,Alford:2004pf}: the constant $c$, which simulates perturbative interactions,
is fixed to a value of $0.05$ for which we obtain a maximum mass for cold and catalyzed hybrid stars
of $1.94 M_{\odot}$, in agreement with the very recent measurement of a compact star having a mass of
$1.97 \pm 0.04 M_{\odot}$ \cite{Demorest:2010bx}.
For this choice of
parameters, the onset of the phase transition occurs at twice nuclear
saturation density for the initial lepton fraction and entropy
configurations.
\begin{figure}
    \begin{centering}
\epsfig{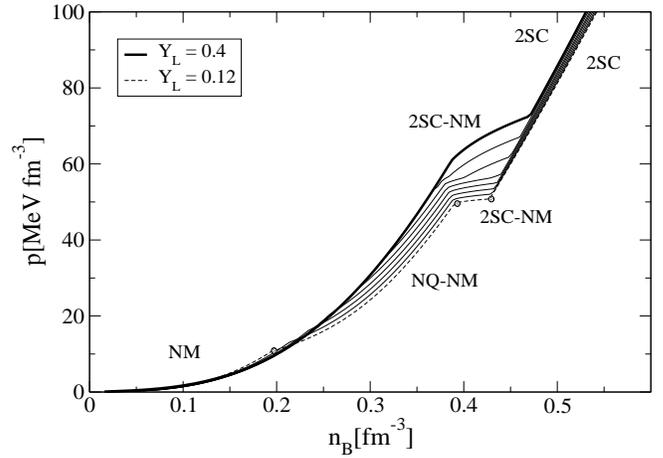}
    \caption{Pressure as a function of the baryon density for the nucleonic matter - quark matter 
phase transition for values of $Y_L$ ranging from $0.4$ (thick line) to $0.12$ (dashed line) with intervals
of $0.04$ (solid lines). The dots on the dashed line indicate the onset and the end 
of the mixed phases of nucleonic matter and normal quark matter and nucleonic matter and 2SC phase. 
At large values of $Y_L$, the quark phase corresponds to the 2SC phase. At lower $Y_L$, the mismatch 
between the chemical potentials of up and down quarks is such that the pairing is broken and 
the normal quark phase is obtained.}
   \end{centering}
\end{figure} 
The phase transition
is computed by using the Gibbs construction as described in
\cite{Prakash:1995uw,Drago:1997tn,Steiner:2000bi,Hempel:2009vp,Pagliara:2010qm} with the conditions
of global conservation of baryonic number, lepton number and electric charge.
The resulting extended mixed phase is obtained by solving the following equations:
\begin{eqnarray}
p^H(\mu_B^H,\mu_C^H,\mu_L^H,T^H) = p^Q(\mu_B^Q,\mu_C^Q,\mu_L^Q,T^Q)\\
(1-\chi)n^H_C + \chi n^Q_C -  n_e=0\\
n_e+n_{\nu}= Y_L n_B\\
(1-\chi)s^H+\chi s^Q= S/Nn_B\\
T^H=T^Q\\
\mu_B^H=\mu_B^Q,\,\, \mu_C^H=\mu_C^Q,\,\,\mu_L^H=\mu_L^Q
\end{eqnarray}
where $p^{i}, \mu_B^i, \mu_C^i, \mu_L^i, T^i, n^i_C, s^i$ are
the pressure, baryon chemical potential, charge chemical potential,
lepton chemical potential, temperature, electric charge density and
entropy density of the hadronic ($i=H$) and the quark phases ($i=Q$);
$n_e$ and $n_{\nu}$ are the electron and electron neutrino densities,
$n_B$ is the baryon density and $\chi$ is the volume fraction of the
quark phase. It has been recently pointed out that another possible
hadron-quark mixed phase could appear in protoneutron stars if local
charge neutrality and global lepton number conservation are imposed
\cite{Pagliara:2009dg}; we will not consider this possibility here.

In Fig.~\ref{fig:pressnucl}, we show the EoS (pressure vs baryon
density) of nucleonic matter for different values of $Y_L$.  Notice
that by decreasing $Y_L$, the EoS becomes stiffer: despite the fact
that at lower $Y_L$ the contribution to the total pressure of leptons is
smaller, the lower values of $Y_e$ implies a larger isospin asymmetry
of nucleonic matter. The effect of the symmetry energy thus dominates
over the reduced leptonic pressure and the EoS is stiffer. This a well
known fact and it also implies that the maximum baryonic mass of
protoneutron stars is smaller than the maximum baryonic mass of cold
neutron star (if nucleonic matter is considered)
\cite{Prakash:1996xs}. 

Let us discuss now our results for the mixed
phase. We show the EoS in Fig.~2: at a large lepton
fraction, $Y_L=0.4$, after a low density nucleonic phase, the 2SC-NM
mixed phase starts and at high density the pure 2SC phase takes place.
By decreasing the value of $Y_L$, at some point, close to the onset of
the mixed phase, the 2SC pairing cannot take place anymore because of
the large mismatch between the up and down chemical potentials (the
isospin density increases as $Y_L$ decreases) and the normal quark
phase starts to form. Notice that the transition from the 2SC phase
to the normal phase occurs close to the onset of the mixed phase and
moves to higher densities as $Y_L$ is further decreased. This is due
to the fact that the local isospin density of the quark phase is large
at the onset of the mixed phase and gradually decreases as the volume
fraction increases, as shown in
Refs.~\cite{DiToro:2006pq,DiToro:2009ig,Pagliara:2010ii}. The largest 
stress on the up and down chemical potentials is thus realized close to the 
onset of the mixed phase: the hadron quark phase transition induces also 
a phase transition between quark phases. 
\begin{figure}
    \begin{centering}
\epsfig{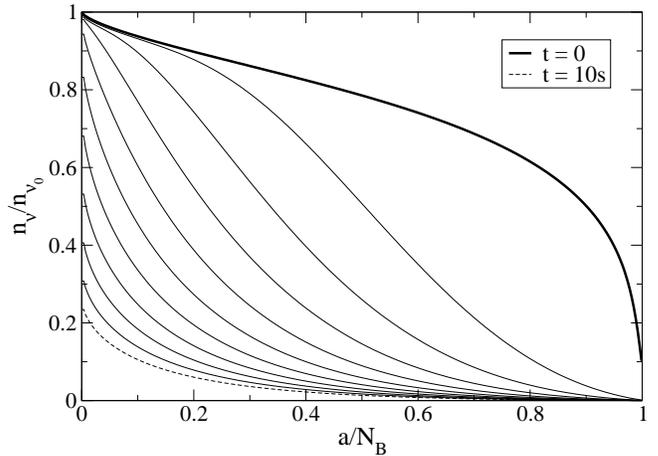}
    \caption{Temporal variation of the 
neutrino density as a function of the enclosed baryonic number for a neutron star 
of total baryonic number $N_B=2.29 \times 10^{57}$. 
The initial central neutrino density at the center of the star, $n_{\nu_0}=1.88\times 10^{-2}$fm$^{-3}$. 
The thick line corresponds to the initial configuration, $t=0$,
and the dashed line to the configuration at $t=10$s. 
The thin solid lines stand for the intermediate configurations
with temporal steps of $1$s.}
   \end{centering}
\end{figure} 
To calculate the transition between the NQ and the 2SC phases within
the mixed phase we use the following simple prescription: we start
with the high density pure 2SC phase, $\chi=1$, and we decrease the density until
we reach the border of the mixed phase with the NM obtained by solving the
system of Eqs.~1-6; we solve the same system at decreasing
values of the baryon density and $\chi$ and obtain the solutions $\mu_B^*$,
$\mu_C^*$, $\mu_L^*$ and $T^*$; then we accept the solution of the
system only if the pressure of the normal quark phase computed at the
same $\mu_B^*$, $\mu_C^*$, $\mu_L^*$ and $T^*$ is lower than the
pressure of the 2SC phase. When this condition is not fulfilled
anymore we choose the NQ phase as the quark component of the mixed
phase. This is a simple way to describe the unpairing phase transition
between the 2SC and the NQ phases (similar to a Maxwell
construction) within the mixed phase with NM. Actually, since this phase transition is also of first
order one should introduce two volume fractions associated with the
two quark phases $\chi_{NQ}$ and $\chi_{2SC}$ and solve a more
complicated system of equations which provides a mixed phase made of
three components. A similar calculation has been done, for cold and beta stable matter, for mixed phases
composed by quark color superconducting phases \cite{Neumann:2002jm},
but calculations for protoneutron star matter, also including the nucleonic phase,
have never been performed. Our procedure, while simple, is sensible since the
NQ-2SC phase transition is not a strong first order, i.e. the jump of
the density is not large when going from one phase to the other one
(the density of the 2SC phase is larger than the one of the NQ by an
additive term scaling as $(\Delta/\mu)^2$, with $\Delta \sim 100$ MeV being the
superconducting gap). Therefore the three components mixed phase would
actually cover a small interval of density. We consider a full
calculation with the three components as an interesting future development of
this work.
\begin{figure}
    \begin{centering}
\epsfig{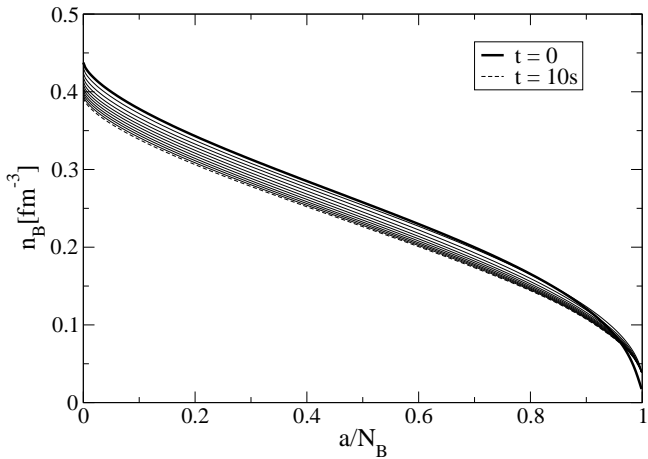}
    \caption{Temporal variation of the 
baryon density as a function of the enclosed baryonic number for a neutron star 
of total baryonic number $N_B=2.29 \times 10^{57}$. 
The thick line corresponds to the initial configuration $t=0$
and the dashed line to the configuration at $t=10$s. The thin solid lines stand for the intermediate configurations
with temporal steps of $1$s. A small decrease of the central density occurs during deleptonization.}
   \end{centering}
\end{figure}

\section{Neutrino diffusion in protoneutron stars}
The evolution of a protoneutron star can be schematically
divided into two separate stages \cite{Prakash:1996xs}: the
deleptonization, during which the trapped neutrinos diffuse and are
gradually released from the star, and the cooling, during which the
entropy of the star reduces by the emission of free streaming
neutrinos. During deleptonization, the star is actually reheated by the
interactions of neutrino with baryonic matter and the initial $S/N
\sim 1$ raises to a value of $\sim 2$.  
This is for us the most interesting stage because
$Y_e$ and $Y_p $ also decrease during it thus leading to a sizable change of
the chemical composition of the star which, as we will show, is responsible for
the formation of different quark phases. The standard technique to
simulate the evolution of protoneutron stars consists of solving
two partial differential equations associated with the transport of
lepton number and the transport of energy within the so called diffusion
approximation
\cite{Burrows:1981zz,Burrows:1986me,Keil:1995hw,Pons:1998mm}. These
equations are coupled with the Tolman-Oppenheimer-Volkoff (TOV) equations to
take into account the mechanical readjustment of the star during the
evolution. Indeed, the time scales of diffusion are much larger than
the dynamical time scale of the compact objects and therefore the
evolution of the structure of the star follows quasi-static
equilibrium configurations. In the cases in which the deleptonization
drives a collapse of the star, because of the formation of exotic
phases for instance, the hydrodynamical evolution must be also taken
into account as done in \cite{Baumgarte:1996iu} or more recently in
\cite{Fischer:2009af,Huedepohl:2009wh} where for the first time
consistent simulations from the core collapse of the supernova to the cooling epoch of the newly formed neutron star
have been performed.
 \begin{figure}
    \begin{centering}
\epsfig{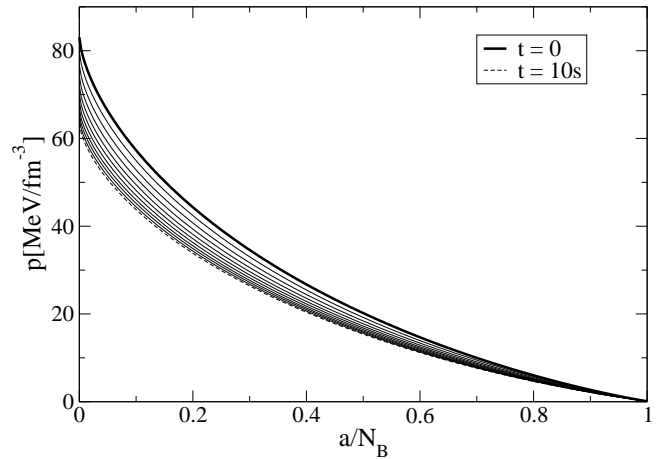}
    \caption{Temporal variation of the 
pressure as a function of the enclosed baryonic number for a neutron star 
of total baryonic number $N_B=2.29 \times 10^{57}$. 
The thick line corresponds to the initial configuration $t=0$
and the dashed line to the configuration at $t=10$s. The thin solid lines stand for the intermediate configurations
with temporal steps of $1$s. A small decrease of the central pressure occurs during deleptonization.
\label{fig:pa-nuclear} }
   \end{centering}
\end{figure} 

The crucial ingredients for such calculations are the EoS and the
cross sections of all the possible interactions of neutrinos with the
matter of the star. Different studies have addressed the possibility
to form ``exotic'' phases during the evolution of a protoneutron star
as hyperons \cite{Keil:1995hw,Pons:1998mm}, kaon condensates
\cite{Pons:2000iy} or finally normal quark matter
\cite{Pons:2001ar}. A similar calculation taking into account color superconducting
phases is still missing and this is the problem that we want to
address in this paper. Since we want here to provide a qualitative description
on how the unpairing transition could occur in a compact star, we will
adopt the following simplifying assumptions in the treatment of the neutrino
transport in protoneutron stars: we will consider only the
first seconds of the stage of deleptonization when neutrinos are still
degenerate and we will assume the entropy profile to be constant
during the evolution ($S/N=1$). These assumptions lead to several
simplifications: we will not solve the equation for the energy
transport but just the one describing the lepton number transport coupled with
the TOV equations; when neutrinos are degenerate the dominant
contribution to the mean free path is given by the absorption
processes of neutrinos by non-degenerate neutrons in NM and by
degenerate quarks in the quark phases
\cite{Prakash:1996xs,Steiner:2001rp}. Moreover, also regarding the
EoS to be used as input for the simulations we will use a
two-dimensional table with the baryon density and the lepton fraction
as independent variables and the ratio $S/N$ will be kept constant.
Because of our simplifying assumptions, our calculation has two important limitations: 
the estimate for the timing of the phase transition must be taken only as an order of
magnitude estimate; indeed, because of the reheating that we are neglecting
here, all the cross sections of neutrinos would be enhanced by
the increase of the temperature and therefore we expect the global
evolution of the system to be slower than what we find here. Secondly,
we cannot provide here an estimate of the neutrino luminosity which is
obtained by solving the energy transport equation. 

Let us discuss now the equations that we solve to simulate the deleptonization \cite{Keil:1995hw}:

\bq
n_B\frac{\partial Y_L}{\partial t}&=&\frac{\Gamma}{r^2}\frac{\partial}{\partial r}(r^2 e^{\phi}\frac{\lambda}{3}\frac{\partial n_{\nu_e}}{\partial r})\\
\frac{dp}{dr}&=& -(p+e)\frac{m+4\pi r^3 p}{r^2-2m r}\\
\frac{dm}{dr}&=& 4 \pi r^2 e \\
\frac{da}{dr}&=& \frac{4 \pi r^2 n_B}{\sqrt{1-2m/r}}\\
\frac{d \phi}{dr}&=& \frac{m+4 \pi r^3 p}{r^2-2 m r}
\eq

where $r$ is the radial distance in spherical coordinates,
$\Gamma=\sqrt{1-2m/r}$, and $e^{\phi}$ are the general relativity
corrections to the diffusion equation, $\lambda$ is the spectral
average of the electron neutrino mean free path, $p$, $e$, $m$ and $a$
are the pressure, the energy density, the enclosed gravitational mass
and the enclosed baryonic number (we use the gravitational units $G=1$
and $c=1$). The first equation describes the diffusion of neutrinos
driven by the neutrino density gradient (i.e. within the diffusion
approximation ) and the last four equations describe the mechanical
equilibrium of the star, the TOV equations. The initial condition and
the boundary conditions are specified as follows: for the initial
condition we use an EoS with constant $S/N=1$ and $Y_L=0.4$; on the
boundaries, $r=0$ and $r=R$ (where $R$ is the radius of the star
which, numerically, corresponds to a very small value of the
pressure), the neutrino flux $F(r)=-\frac{\lambda}{3}\frac{\partial
n_{\nu_e}}{\partial r}$ is taken as $F(0)=0$ and $F(R)=f n_{\nu_e}$
where $f$ is a constant varied between $0.1-0.6$. The boundary
condition at $r=R$ basically imposes that the flux of neutrinos is a
certain fraction $f$ of the flux that one would have if neutrinos were
free streaming \cite{Pons:1998mm}. We also need initial conditions for
the TOV equations: the central pressure is chosen arbitrarily,
$p(r=0)=p_c$ (different values of $p_c$ lead to different values for
the total baryon number of the star), 
$m(r=0)=0$,
$a(r=0)=0$ and $\phi(r=R)=\frac{1}{2}log(1-2m/r)$.
\begin{figure}
    \begin{centering}
\epsfig{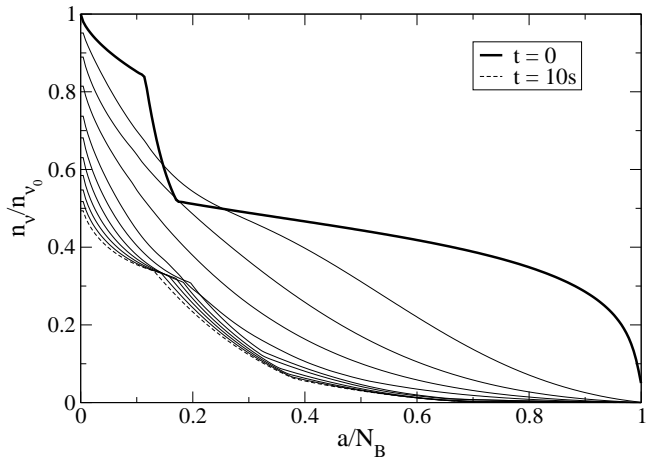}
    \caption{Temporal variation of the 
neutrino density as a function of the enclosed baryonic number for a hybrid star of baryonic number $N_B=2.29 \times 10^{57}$. 
The initial central neutrino density, $n_{\nu_0}=3.44\times 10^{-2}$fm$^{-3}$. The thick line corresponds to the initial configuration $t=0$
and the dashed line to the configuration at $t=10$s. The thin solid lines stand for the intermediate configurations
with temporal steps of $1$s.
\label{fig:nua-nuclear} }
   \end{centering}
\end{figure} 
Finally, concerning the mean free paths,
during the initial stage of the evolution, the absorption processes
are the dominant ones and thus we take for NM the mean
free path associated with the absorption of degenerate neutrinos by non-degenerate
neutrons \cite{Prakash:1996xs}:
\begin{equation} 
\lambda_n=\frac{4}{n_n \sigma_0 (1+3 g_A^2)}\left(\frac{m_e}{E_{\nu}}\right)^2
\end{equation}
where $n_n$ is the neutron number density, $\sigma_0=1.76 \times
10^{-44}$ cm$^2$, $g_A=1.257$, $m_e$ is the electron mass and $E_{\nu}$,
the neutrino energy.  This formula must be actually corrected by a
factor of $3-10$ to take into account the degeneracy and Fermi liquid
corrections \cite{Prakash:1996xs}, we will use a factor of ten in this paper.  
Similarly for quark matter we
consider the absorption processes of degenerate neutrinos by
degenerate down quarks \cite{Steiner:2001rp}:
\begin{equation} 
 \lambda_q=\frac{5 \pi^3 \mu^2_{\nu}}{2 G_F^2 \mu_e^3(10\mu_u^2+5\mu_u\mu_e+\mu_e^2)((E_{\nu}-\mu{_\nu})^2+\pi^2T^2)}
\end{equation}
This formula holds for normal quark matter and it can be used also for
the blue quarks of the 2SC phase which are indeed unpaired. We
therefore multiply $\lambda_q$ by a factor of three within the 2SC
phase to take into account that the absorption process can take place
only if blue quarks are involved. For the paired red and green quarks
of the 2SC phase, the corresponding mean free path is larger
\cite{Carter:2000xf} then the one associated with the blue free quarks and
thus we can safely neglect them in our calculation. Finally, since we
study the regime in which neutrinos are still degenerate, the spectral
average of the mean free path is dominated by neutrinos at the Fermi
surface \cite{Burrows:1981zz} thus we set, in Eq.~12-13, $E_{\nu}=\mu_{\nu}$.  A last
comment about the neutrino mean free path within the mixed phase
$\lambda_{MP}$: it is simply a weighted sum of the mean free paths of
the two phases: $1/\lambda_{MP}=\chi/\lambda_q+(1-\chi)/\lambda_n$.
\begin{figure}
    \begin{centering}
\epsfig{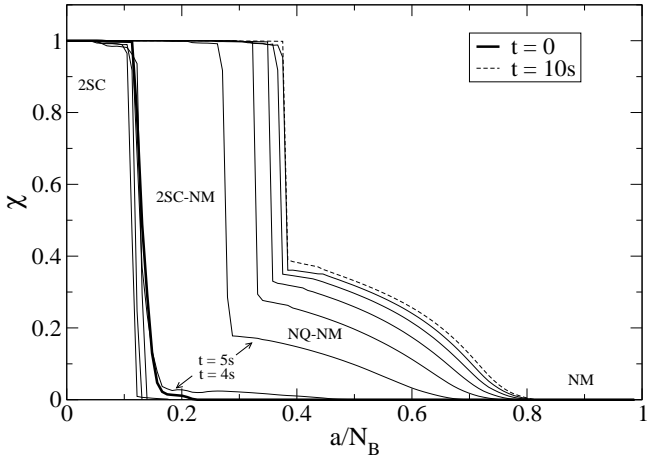}
    \caption{Temporal evolution of the volume fraction of the quark phase for a hybrid star of baryonic number $N_B=2.29 \times 10^{57}$. Initially 
the star has a core of 2SC phase, a thin layer of mixed phase and an outer layer of nucleonic matter. After $\sim 4$ s a sizable change
of the structure of the star occurs due to the formation of the normal quark phase triggered by deleptonization. 
The mixed phase has three components: nucleonic matter, 2SC and normal quark matter. }
   \end{centering}
\end{figure} 

Let us discuss now our results by starting from the case of
the deleptonization of an ordinary neutron star, which will constitute
our reference model. We fix the central pressure of the star $p_c=83$
MeV/fm$^3$, and calculate the equilibrium configuration for the
initial uniform profiles of $S/N$ and $Y_L$.  We stop the integration of the
structure equations when the pressure drops to a value $p_0$
corresponding to a baryon density of roughly half of the nuclear
saturation density \footnote{At lower densities, where the
crust takes place, the diffusion approximation does not hold anymore
because neutrinos are free streaming and a more sophisticated
transport calculation would be needed. Here we want to study only the
change of the chemical composition of the core of the star thus we can
neglect the low density part of the star.}. We obtain a baryon number of $N_B \sim 2.29
\times 10^{57}$ with a corresponding gravitational mass of $\sim 1.78M_{\odot}$.  
We then start the diffusion simulation \footnote{We use,
for solving Eqs.~7-11, an iterative method also used in
\cite{Pons:1998mm}. We solve at a certain time step Eq.~7 by using an
implicit finite difference method (in a time - enclosed baryonic number
grid ) and we insert the obtained profile of $Y_L$ in the structure
equations; we calculate then the equilibrium configuration by using
the Runge-Kutta method and we plug back the adjusted values of the
thermodynamical and stellar quantities (baryon density, pressure,
gravitational mass, radius, etc.) in Eq.7 and we re-calculate its
solution for the same temporal step. We repeat this procedure until
convergence is reached. An important point in this calculation is the
conservation of the total baryon number. This constraint is fulfilled,
when solving the structure equations, by changing the value of the
central pressure of the star until we obtain $a(p_0)=N_B$ at the
last Runge-Kutta iteration.}.  Results for the temporal variations of
the neutrino density, the baryon density and the pressure profiles are
shown in Figs.~3-5. From Fig.~3 one can notice that while we are here
adopting many simplifying assumptions for the transport process, the
time scale of deleptonization is in the right order of magnitude:
after $t^*\sim 7$s the central neutrino density is half of its initial
value.  Similar estimates are obtained with the simple treatment of
the deleptonization and cooling time scales presented in
\cite{Prakash:1996xs} and in the simulations of \cite{Burrows:1986me} (on the other hand, in
the more recent one \cite{Pons:1998mm}, a better treatment of the
baryon - neutrino interactions leads actually to a deleptonization
process which continues through most of the cooling epoch, $t_{del}
\sim 50$s). We stop our simulation at $t=10$s, since for longer times
neutrino would start to become non-degenerate and our assumptions do
not hold anymore.

Instead, the temporal behavior of the density and the pressure is
different from what was found in \cite{Burrows:1986me,Pons:1998mm} but in
qualitative agreement with the results of \cite{Keil:1995hw} (see Figs.~3 and 10 of that paper): during
the first seconds of the evolution, while in the outer layers  of the star ($a/N_B$
close to $1$) the density increases due to the lack of
neutrino pressure, the pressure and density at the center of the star
slightly decrease. This is actually a consequence of the stiffening of
the nucleonic EoS when $Y_L$ is reduced: even if the contribution to
the total pressure of neutrinos decreases during deleptonization,
nucleonic matter becomes more and more isospin asymmetric and, due to
the large value of the symmetry energy at high density within the TM1
model (see \cite{Pagliara:2010ii}), the overall effect is a stiffer
EoS. This effect deserves more accurate studies by implementing a different choice
for the initial profiles of $Y_L$ and $S/N$ taken from supernova
simulations and by improving the treatment of the neutrino and energy
transport (energy transport is not included here). Interestingly, in the recent sophisticated calculation presented in
Ref.~\cite{Fischer:2009af}, the central density is almost constant during the long term evolution (see Fig.~16a of that paper) 
at variance with the results of \cite{Pons:1998mm}. 
Here we want just to stress that the symmetry energy could 
play an important role in protoneutron star evolution and, since quark matter has usually a smaller symmetry energy than 
nucleonic matter, hybrid stars show a different temporal evolution as we will show
in the following. 

By looking at the
pressure profiles in Fig.~5, we notice that the value of the pressure
at $a=N_B$ is the same at all times as it should be, which represents an
important check of our calculation.

\begin{figure}
    \begin{centering}
\epsfig{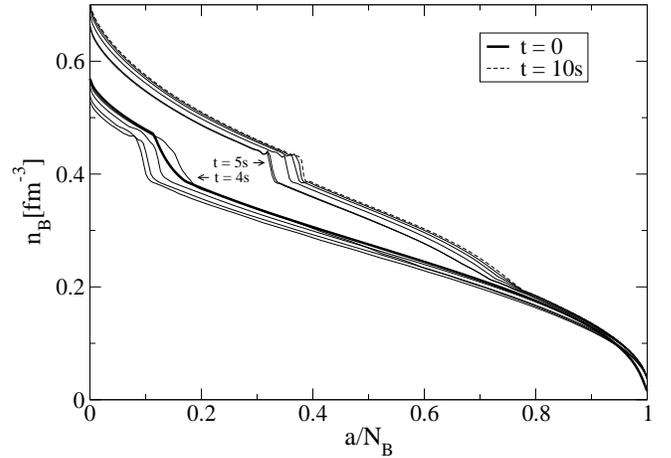}
    \caption{Temporal variation of the 
baryon density as a function of the enclosed baryonic number for a hybrid star of baryonic number $N_B=2.29 \times 10^{57}$. 
The thick line corresponds to the initial configuration $t=0$
and the dashed line to the configuration at $t=10$s. The thin solid lines stand for the intermediate configurations
with temporal steps of $1$s. The central density slightly decreases during the first seconds but
after $\sim 4$s when the normal quark phase appears, it starts to increase.}
   \end{centering}
\end{figure}

Let us discuss now the results obtained for the hybrid EoS presented
in Sec.~II. We consider an hybrid star having the same baryon number as
the neutron star considered before. The corresponding central pressure
is $p_c=118$ MeV/fm$^3$ and the initial gravitational mass $M = 1.78
M_{\odot}$. After calculating the initial configuration with constant
$S/N$ and $Y_L$, we start the temporal evolution simulation. The
temporal variation of the neutrino density is shown in Fig.~6: notice that the neutrino density profile follows
the same shape of the baryon density profile (Fig.~8), with the typical
features of a first order phase transition (a mixed phase and two pure
phases). In the pure 2SC quark phase, the density of neutrinos is
larger than in the pure nucleonic phase, the central neutrino density
is in this case $n_{\nu_0}=3.44\times 10^{-2}$fm$^{-3}$.  This is due
to the depletion of electrons in the quark phase, with respect to the
nucleonic phase, and the condition of fixed lepton fraction
\cite{Pons:2001ar}.  The deleptonization time of the hybrid star is
comparable with the one of the neutron star: although the mean free
path within the quark phase is larger than the one in the nucleonic
phase at the same value of baryon density \cite{Steiner:2001rp},
neutrinos enter the nucleonic phase after diffusing inside the quark phase
and the mixed phase. Since they are more abundant than in the case of
the neutron star, their mean free path within the nucleonic phase is
strongly suppressed (see the dependence of $\lambda_n$ on $E_{\nu}$
in Eq.~(12)).
The modification of the structure
and the composition of the star is shown in Fig.~7 where the quark
volume fraction profile is displayed. This figure represents the main
result of this paper. Starting with an initial configuration with a
core of 2SC phase, a small region of mixed phase and a nucleonic
matter layer, the star gradually modifies its structure and
composition during deleptonization, and at some point ($t \sim 4$s in
this example), close to the onset of the mixed phase, the 2SC pairing
pattern is broken and the normal quark phase is formed. The mixed
phase has now three components and occupies a much larger volume
fraction of the star and continues to grow until a stationary
configuration is reached when almost all neutrinos leak out of the
star. Quite remarkably, in a short amount of time, between $4$ and $5$ s
the structure of the star changes sizably with possible observable effects. 
This change of the structure of the star is also evident
from the baryon density profile shown in Fig.~8.
The baryon density slightly decreases during the
first seconds, similarly to the neutron star case, but after $\sim 4$s,
in coincidence with the formation of the normal quark phase in the
mixed phase, it starts to increase. Initially only a
small fraction of the star is occupied by the 2SC phase and therefore
the evolution of the system is dominated by the nucleonic phase. With
the formation of the normal quark phase, very quickly a large
fraction of the star is occupied by the quark and the mixed phase and
the density starts to increase further favoring the formation of the
quark phase. This behavior is compatible with the fact that the EoS of the
mixed phase and the pure quark phase softens as the lepton fraction
decreases (see Fig.~2.). This qualitative difference between neutron
stars and stars containing an exotic phase was already found in
\cite{Keil:1995hw} for the case of hyperonic matter. Although
very interesting, because related with the different symmetry energy of
the nucleonic phase and the quark phase, the different evolution of
the density in neutron stars and hybrid stars must be investigated in
more sophisticated transport models and by using more realistic
initial conditions before a firm conclusion can be drawn. In the same
figure, one can also notice how the profile of the density
qualitatively changes during the evolution because of the sizable
modification of the mixed phase. 
Unfortunately, we cannot provide here signatures of this
structural and chemical changes in the star within the neutrino signal
released; such a calculation represents an important extension of this
work.

Let us consider one more case with a smaller protoneutron star total
baryon number. Fig.~9 shows the temporal evolution of the neutrino
density for a neutron star having an initial gravitational mass of
$1.38 M_{\odot}$ and $N_B=1.74 \times 10^{57}$. The initial central
density is smaller than the density of the onset of the mixed phase,
the evolution proceeds as in the case of neutron stars with a slow
decrease of the central density and after $t \sim 6$s, when a large
part of neutrinos already left the star, the
normal quark phase starts to appear in the mixed phase and reaches an
equilibrium volume fraction of $\sim 0.2$, shown in the insert of Fig.~9
(this scenario is similar to the cases considered in
\cite{Pons:2001ar}).  In this case, the 2SC phase cannot be formed
because the initial conditions are such that quark matter appears when
matter is already very isospin asymmetric.

For very large initial total baryon number, a collapse is obtained
during the evolution because the maximum mass is reached. The scenario
in that case is quite similar to the one obtained for hyperonic
matter, kaon condensed matter and normal quark matter
\cite{Keil:1995hw,Baumgarte:1996iu,Pons:1998mm,Pons:2000iy,Pons:2001ar}.
Due to our simple treatment of the initial conditions and the neutrino
transport we cannot provide here the evolution of configurations close
to the maximum mass since for those cases the initial conditions are
crucial for the stability of the star with respect to collapse.  A
full calculation from the supernova explosion to the late evolution of
protoneutron stars is needed for very massive stellar objects as done
in \cite{Fischer:2009af,Huedepohl:2009wh}.

\begin{figure}
    \begin{centering}
\epsfig{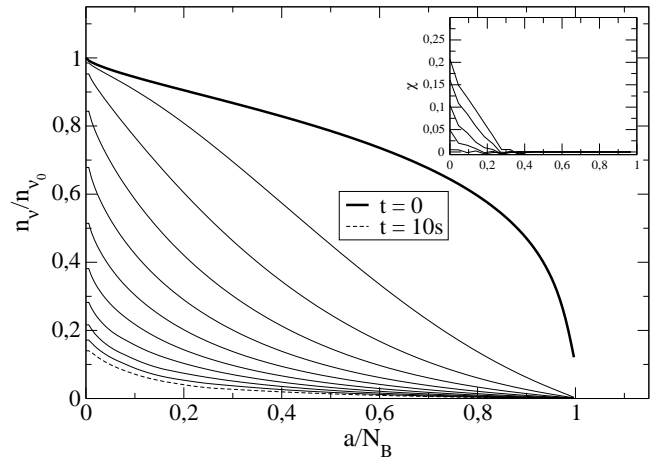}
    \caption{Temporal variation of the 
neutrino density as a function of the enclosed baryonic number for a hybrid star of baryonic number $N_B=1.74 \times 10^{57}$. 
The initial central neutrino density, $n_{\nu_0}=1.66 \times 10^{-2}$fm$^{-3}$. The thick line corresponds to the initial configuration $t=0$
and the dashed line to the configuration at $t=10$s. The thin solid lines stand for the intermediate configurations
with temporal steps of $1$s. The insert shows the quark volume fraction profiles for the latest $4$ seconds of the evolution. 
Only the normal quark phase
is formed in this case at late stages of the deleptonization process.}
   \end{centering}
\end{figure}


\section{Conclusions}
We have computed the temporal evolution of protoneutron stars
containing a color superconducting phase in the core. We considered
here the case of the 2SC phase which, between the different color
superconducting phases, is the most likely to appear in a newly born
neutron star. During the first few seconds after the birth of the
star, the gradual deleptonization is accompanied by a reduction of the
electron and proton fractions: matter becomes more and more isospin
asymmetric. In the 2SC quark phase, when the isospin asymmetry reaches
a certain threshold the pairing cannot take place anymore and a first
order phase transition to the normal quark phase
occurs. Interestingly, the unpairing transition takes place close to
the onset of the nucleonic matter - quark matter mixed phase, where the
mismatch between the quark chemical potentials is the highest.  The
structure and composition of the star are strongly modified during the
phase transition and they reach an equilibrium configuration at late
times when most of the neutrinos are released.  This is the typical
evolution of a star with initial gravitational mass larger than $\sim
1.6 M_{\odot}$.  For less massive hybrid stars, quark matter appears
only at later times and the isospin asymmetry is then so large that
the 2SC phase cannot be formed and only a normal quark phase can appear
in the core of the star.

The calculations we have presented here are based on many
simplifying assumptions for the neutrino interactions and for the
treatment of the neutrino transport. We have nevertheless provided a few
examples of how the evolution of a protoneutron star can be rich and
complex when color superconducting phases are taken into account.
More realistic calculations are of course needed and also the three
flavor CFL phase should be included in the equation of state. Work
along this line is already in progress.

Finally we want to remark why the scenario we have discussed here is
potentially very interesting for phenomenology: the neutrino signal,
could be actually modified by the happenings of the core of the star
as shown in \cite{Sagert:2008ka,Dasgupta:2009yj,Fischer:2010wp} for the scenario of a
phase transition to quark matter in the early post-bounce stage of a
supernova. If sizable modifications of the structure of the star
occurs one could indeed expect to have signatures on the spectrum and
the luminosity of the emitted neutrinos: for instance, the average
energy of the emitted neutrinos strongly depends on the compactness
of the star \cite{Fischer:2009af}. Also the effect of metastability
and nucleation connected with a first order phase transition
\cite{Mintz:2009ay} could have an imprint on the temporal structure
of the signal. The present neutrino detectors, as SuperK, will
be able to catch thousands of neutrinos for a galactic supernova thus
allowing a detailed temporal and spectral investigation of the signal
\cite{Dasgupta:2009yj}.

Another very exciting implication of our scenario concerns
gamma-ray-bursts: there are now many hints on the possibility that at
least some gamma-ray-bursts are produced by strongly magnetized
neutron stars.  The magnetar model of gamma-ray-bursts
\cite{Metzger:2010tn} as compared to the collapsar model could explain
better the long term activity, up to $10^6$ s after the prompt
emission, seen in some light curves
\cite{Dall'Osso:2010ah,Bernardini:2010zb}. This implies that the
prompt emission is connected with the first stages of the evolution of
a protoneutron star and the specific features of the neutrino signal
could have therefore an impact on the gamma-ray-bursts.  The complex
temporal structure of some gamma-ray-bursts could have its explanation
in the complex phenomena which can occur in a protoneutron star if
quark phases are formed.

\bigskip
This work is supported by the Deutsche Forschungsgemeinschaft (DFG)
under Grant No. PA 1780/2-1. The author thanks J. Schaffner-Bielich
for valuable discussions.  This work was also supported by CompStar, a
Research Networking Programme of the European Science Foundation.


\begin{thebibliography}{51}
\expandafter\ifx\csname natexlab\endcsname\relax\def\natexlab#1{#1}\fi
\expandafter\ifx\csname bibnamefont\endcsname\relax
  \def\bibnamefont#1{#1}\fi
\expandafter\ifx\csname bibfnamefont\endcsname\relax
  \def\bibfnamefont#1{#1}\fi
\expandafter\ifx\csname citenamefont\endcsname\relax
  \def\citenamefont#1{#1}\fi
\expandafter\ifx\csname url\endcsname\relax
  \def\url#1{\texttt{#1}}\fi
\expandafter\ifx\csname urlprefix\endcsname\relax\def\urlprefix{URL }\fi
\providecommand{\bibinfo}[2]{#2}
\providecommand{\eprint}[2][]{\url{#2}}

\bibitem[{\citenamefont{Alford et~al.}(2008)\citenamefont{Alford, Schmitt,
  Rajagopal, and Schafer}}]{Alford:2007xm}
\bibinfo{author}{\bibfnamefont{M.~G.} \bibnamefont{Alford}},
  \bibinfo{author}{\bibfnamefont{A.}~\bibnamefont{Schmitt}},
  \bibinfo{author}{\bibfnamefont{K.}~\bibnamefont{Rajagopal}},
  \bibnamefont{and} \bibinfo{author}{\bibfnamefont{T.}~\bibnamefont{Schafer}},
  \bibinfo{journal}{Rev. Mod. Phys.} \textbf{\bibinfo{volume}{80}},
  \bibinfo{pages}{1455} (\bibinfo{year}{2008}), \eprint{0709.4635}.

\bibitem[{\citenamefont{Steiner et~al.}(2002)\citenamefont{Steiner, Reddy, and
  Prakash}}]{Steiner:2002gx}
\bibinfo{author}{\bibfnamefont{A.~W.} \bibnamefont{Steiner}},
  \bibinfo{author}{\bibfnamefont{S.}~\bibnamefont{Reddy}}, \bibnamefont{and}
  \bibinfo{author}{\bibfnamefont{M.}~\bibnamefont{Prakash}},
  \bibinfo{journal}{Phys. Rev.} \textbf{\bibinfo{volume}{D66}},
  \bibinfo{pages}{094007} (\bibinfo{year}{2002}), \eprint{hep-ph/0205201}.

\bibitem[{\citenamefont{Ruester et~al.}(2005)\citenamefont{Ruester, Werth,
  Buballa, Shovkovy, and Rischke}}]{Ruester:2005jc}
\bibinfo{author}{\bibfnamefont{S.~B.} \bibnamefont{Ruester}},
  \bibinfo{author}{\bibfnamefont{V.}~\bibnamefont{Werth}},
  \bibinfo{author}{\bibfnamefont{M.}~\bibnamefont{Buballa}},
  \bibinfo{author}{\bibfnamefont{I.~A.} \bibnamefont{Shovkovy}},
  \bibnamefont{and} \bibinfo{author}{\bibfnamefont{D.~H.}
  \bibnamefont{Rischke}}, \bibinfo{journal}{Phys. Rev.}
  \textbf{\bibinfo{volume}{D72}}, \bibinfo{pages}{034004}
  (\bibinfo{year}{2005}), \eprint{hep-ph/0503184}.

\bibitem[{\citenamefont{Blaschke et~al.}(2005)\citenamefont{Blaschke,
  Fredriksson, Grigorian, Oztas, and Sandin}}]{Blaschke:2005uj}
\bibinfo{author}{\bibfnamefont{D.}~\bibnamefont{Blaschke}},
  \bibinfo{author}{\bibfnamefont{S.}~\bibnamefont{Fredriksson}},
  \bibinfo{author}{\bibfnamefont{H.}~\bibnamefont{Grigorian}},
  \bibinfo{author}{\bibfnamefont{A.~M.} \bibnamefont{Oztas}}, \bibnamefont{and}
  \bibinfo{author}{\bibfnamefont{F.}~\bibnamefont{Sandin}},
  \bibinfo{journal}{Phys. Rev.} \textbf{\bibinfo{volume}{D72}},
  \bibinfo{pages}{065020} (\bibinfo{year}{2005}), \eprint{hep-ph/0503194}.

\bibitem[{\citenamefont{Warringa}(2006)}]{Warringa:2006dk}
\bibinfo{author}{\bibfnamefont{H.~J.} \bibnamefont{Warringa}}
  (\bibinfo{year}{2006}), \eprint{hep-ph/0606063}.

\bibitem[{\citenamefont{Casalbuoni et~al.}(2005)\citenamefont{Casalbuoni,
  Gatto, Ippolito, Nardulli, and Ruggieri}}]{Casalbuoni:2005zp}
\bibinfo{author}{\bibfnamefont{R.}~\bibnamefont{Casalbuoni}},
  \bibinfo{author}{\bibfnamefont{R.}~\bibnamefont{Gatto}},
  \bibinfo{author}{\bibfnamefont{N.}~\bibnamefont{Ippolito}},
  \bibinfo{author}{\bibfnamefont{G.}~\bibnamefont{Nardulli}}, \bibnamefont{and}
  \bibinfo{author}{\bibfnamefont{M.}~\bibnamefont{Ruggieri}},
  \bibinfo{journal}{Phys. Lett.} \textbf{\bibinfo{volume}{B627}},
  \bibinfo{pages}{89} (\bibinfo{year}{2005}), \eprint{hep-ph/0507247}.

\bibitem[{\citenamefont{Ruester et~al.}(2006)\citenamefont{Ruester, Werth,
  Buballa, Shovkovy, and Rischke}}]{Ruester:2005ib}
\bibinfo{author}{\bibfnamefont{S.~B.} \bibnamefont{Ruester}},
  \bibinfo{author}{\bibfnamefont{V.}~\bibnamefont{Werth}},
  \bibinfo{author}{\bibfnamefont{M.}~\bibnamefont{Buballa}},
  \bibinfo{author}{\bibfnamefont{I.~A.} \bibnamefont{Shovkovy}},
  \bibnamefont{and} \bibinfo{author}{\bibfnamefont{D.~H.}
  \bibnamefont{Rischke}}, \bibinfo{journal}{Phys. Rev.}
  \textbf{\bibinfo{volume}{D73}}, \bibinfo{pages}{034025}
  (\bibinfo{year}{2006}), \eprint{hep-ph/0509073}.

\bibitem[{\citenamefont{Sandin and Blaschke}(2007)}]{Sandin:2007zr}
\bibinfo{author}{\bibfnamefont{F.}~\bibnamefont{Sandin}} \bibnamefont{and}
  \bibinfo{author}{\bibfnamefont{D.}~\bibnamefont{Blaschke}},
  \bibinfo{journal}{Phys. Rev.} \textbf{\bibinfo{volume}{D75}},
  \bibinfo{pages}{125013} (\bibinfo{year}{2007}), \eprint{astro-ph/0701772}.

\bibitem[{\citenamefont{Lugones et~al.}(2010)\citenamefont{Lugones, do~Carmo,
  Grunfeld, and Scoccola}}]{Lugones:2010gj}
\bibinfo{author}{\bibfnamefont{G.}~\bibnamefont{Lugones}},
  \bibinfo{author}{\bibfnamefont{T.~A.~S.} \bibnamefont{do~Carmo}},
  \bibinfo{author}{\bibfnamefont{A.~G.} \bibnamefont{Grunfeld}},
  \bibnamefont{and} \bibinfo{author}{\bibfnamefont{N.~N.}
  \bibnamefont{Scoccola}}, \bibinfo{journal}{Phys. Rev.}
  \textbf{\bibinfo{volume}{D81}}, \bibinfo{pages}{085012}
  (\bibinfo{year}{2010}), \eprint{1001.1709}.

\bibitem[{\citenamefont{Drago et~al.}(2004)\citenamefont{Drago, Lavagno, and
  Pagliara}}]{Drago:2004vu}
\bibinfo{author}{\bibfnamefont{A.}~\bibnamefont{Drago}},
  \bibinfo{author}{\bibfnamefont{A.}~\bibnamefont{Lavagno}}, \bibnamefont{and}
  \bibinfo{author}{\bibfnamefont{G.}~\bibnamefont{Pagliara}},
  \bibinfo{journal}{Phys. Rev.} \textbf{\bibinfo{volume}{D69}},
  \bibinfo{pages}{057505} (\bibinfo{year}{2004}), \eprint{nucl-th/0401052}.

\bibitem[{\citenamefont{Drago and Pagliara}(2007)}]{Drago:2005rc}
\bibinfo{author}{\bibfnamefont{A.}~\bibnamefont{Drago}} \bibnamefont{and}
  \bibinfo{author}{\bibfnamefont{G.}~\bibnamefont{Pagliara}},
  \bibinfo{journal}{Astrophys. J.} \textbf{\bibinfo{volume}{665}},
  \bibinfo{pages}{1227} (\bibinfo{year}{2007}), \eprint{astro-ph/0512602}.

\bibitem[{\citenamefont{Pagliara and
  Schaffner-Bielich}(2008)}]{Pagliara:2007ph}
\bibinfo{author}{\bibfnamefont{G.}~\bibnamefont{Pagliara}} \bibnamefont{and}
  \bibinfo{author}{\bibfnamefont{J.}~\bibnamefont{Schaffner-Bielich}},
  \bibinfo{journal}{Phys. Rev.} \textbf{\bibinfo{volume}{D77}},
  \bibinfo{pages}{063004} (\bibinfo{year}{2008}), \eprint{0711.1119}.

\bibitem[{\citenamefont{Drago et~al.}(2008)\citenamefont{Drago, Pagliara,
  Pagliaroli, Villante, and Vissani}}]{Drago:2008tb}
\bibinfo{author}{\bibfnamefont{A.}~\bibnamefont{Drago}},
  \bibinfo{author}{\bibfnamefont{G.}~\bibnamefont{Pagliara}},
  \bibinfo{author}{\bibfnamefont{G.}~\bibnamefont{Pagliaroli}},
  \bibinfo{author}{\bibfnamefont{F.~L.} \bibnamefont{Villante}},
  \bibnamefont{and} \bibinfo{author}{\bibfnamefont{F.}~\bibnamefont{Vissani}},
  \bibinfo{journal}{AIP Conf. Proc.} \textbf{\bibinfo{volume}{1056}},
  \bibinfo{pages}{256} (\bibinfo{year}{2008}), \eprint{0809.0518}.

\bibitem[{\citenamefont{Carter and Reddy}(2000)}]{Carter:2000xf}
\bibinfo{author}{\bibfnamefont{G.~W.} \bibnamefont{Carter}} \bibnamefont{and}
  \bibinfo{author}{\bibfnamefont{S.}~\bibnamefont{Reddy}},
  \bibinfo{journal}{Phys. Rev.} \textbf{\bibinfo{volume}{D62}},
  \bibinfo{pages}{103002} (\bibinfo{year}{2000}), \eprint{hep-ph/0005228}.

\bibitem[{\citenamefont{Pagliara and
  Schaffner-Bielich}(2010)}]{Pagliara:2010ii}
\bibinfo{author}{\bibfnamefont{G.}~\bibnamefont{Pagliara}} \bibnamefont{and}
  \bibinfo{author}{\bibfnamefont{J.}~\bibnamefont{Schaffner-Bielich}},
  \bibinfo{journal}{Phys. Rev.} \textbf{\bibinfo{volume}{D81}},
  \bibinfo{pages}{094024} (\bibinfo{year}{2010}), \eprint{1003.1017}.

\bibitem[{\citenamefont{Bedaque}(2002)}]{Bedaque:1999nu}
\bibinfo{author}{\bibfnamefont{P.~F.} \bibnamefont{Bedaque}},
  \bibinfo{journal}{Nucl. Phys.} \textbf{\bibinfo{volume}{A697}},
  \bibinfo{pages}{569} (\bibinfo{year}{2002}), \eprint{hep-ph/9910247}.

\bibitem[{\citenamefont{Kiriyama et~al.}(2001)\citenamefont{Kiriyama, Yasui,
  and Toki}}]{Kiriyama:2001ud}
\bibinfo{author}{\bibfnamefont{O.}~\bibnamefont{Kiriyama}},
  \bibinfo{author}{\bibfnamefont{S.}~\bibnamefont{Yasui}}, \bibnamefont{and}
  \bibinfo{author}{\bibfnamefont{H.}~\bibnamefont{Toki}},
  \bibinfo{journal}{Int. J. Mod. Phys.} \textbf{\bibinfo{volume}{E10}},
  \bibinfo{pages}{501} (\bibinfo{year}{2001}), \eprint{hep-ph/0105170}.

\bibitem[{\citenamefont{Lawley et~al.}(2006)\citenamefont{Lawley, Bentz, and
  Thomas}}]{Lawley:2005ru}
\bibinfo{author}{\bibfnamefont{S.}~\bibnamefont{Lawley}},
  \bibinfo{author}{\bibfnamefont{W.}~\bibnamefont{Bentz}}, \bibnamefont{and}
  \bibinfo{author}{\bibfnamefont{A.~W.} \bibnamefont{Thomas}},
  \bibinfo{journal}{Phys. Lett.} \textbf{\bibinfo{volume}{B632}},
  \bibinfo{pages}{495} (\bibinfo{year}{2006}), \eprint{nucl-th/0504020}.

\bibitem[{\citenamefont{{Partridge} et~al.}(2006)\citenamefont{{Partridge},
  {Li}, {Kamar}, {Liao}, and {Hulet}}}]{2006Sci...311..503P}
\bibinfo{author}{\bibfnamefont{G.~B.} \bibnamefont{{Partridge}}},
  \bibinfo{author}{\bibfnamefont{W.}~\bibnamefont{{Li}}},
  \bibinfo{author}{\bibfnamefont{R.~I.} \bibnamefont{{Kamar}}},
  \bibinfo{author}{\bibfnamefont{Y.}~\bibnamefont{{Liao}}}, \bibnamefont{and}
  \bibinfo{author}{\bibfnamefont{R.~G.} \bibnamefont{{Hulet}}},
  \bibinfo{journal}{Science} \textbf{\bibinfo{volume}{311}},
  \bibinfo{pages}{503} (\bibinfo{year}{2006}), \eprint{arXiv:cond-mat/0511752}.

\bibitem[{\citenamefont{Bedaque et~al.}(2003)\citenamefont{Bedaque, Caldas, and
  Rupak}}]{Bedaque:2003hi}
\bibinfo{author}{\bibfnamefont{P.~F.} \bibnamefont{Bedaque}},
  \bibinfo{author}{\bibfnamefont{H.}~\bibnamefont{Caldas}}, \bibnamefont{and}
  \bibinfo{author}{\bibfnamefont{G.}~\bibnamefont{Rupak}},
  \bibinfo{journal}{Phys. Rev. Lett.} \textbf{\bibinfo{volume}{91}},
  \bibinfo{pages}{247002} (\bibinfo{year}{2003}), \eprint{cond-mat/0306694}.

\bibitem[{\citenamefont{Prakash et~al.}(1995)\citenamefont{Prakash, Cooke, and
  Lattimer}}]{Prakash:1995uw}
\bibinfo{author}{\bibfnamefont{M.}~\bibnamefont{Prakash}},
  \bibinfo{author}{\bibfnamefont{J.~R.} \bibnamefont{Cooke}}, \bibnamefont{and}
  \bibinfo{author}{\bibfnamefont{J.~M.} \bibnamefont{Lattimer}},
  \bibinfo{journal}{Phys. Rev.} \textbf{\bibinfo{volume}{D52}},
  \bibinfo{pages}{661} (\bibinfo{year}{1995}).

\bibitem[{\citenamefont{Steiner et~al.}(2000)\citenamefont{Steiner, Prakash,
  and Lattimer}}]{Steiner:2000bi}
\bibinfo{author}{\bibfnamefont{A.}~\bibnamefont{Steiner}},
  \bibinfo{author}{\bibfnamefont{M.}~\bibnamefont{Prakash}}, \bibnamefont{and}
  \bibinfo{author}{\bibfnamefont{J.~M.} \bibnamefont{Lattimer}},
  \bibinfo{journal}{Phys. Lett.} \textbf{\bibinfo{volume}{B486}},
  \bibinfo{pages}{239} (\bibinfo{year}{2000}), \eprint{nucl-th/0003066}.

\bibitem[{\citenamefont{Shen et~al.}(1998)\citenamefont{Shen, Toki, Oyamatsu,
  and Sumiyoshi}}]{Shen:1998gq}
\bibinfo{author}{\bibfnamefont{H.}~\bibnamefont{Shen}},
  \bibinfo{author}{\bibfnamefont{H.}~\bibnamefont{Toki}},
  \bibinfo{author}{\bibfnamefont{K.}~\bibnamefont{Oyamatsu}}, \bibnamefont{and}
  \bibinfo{author}{\bibfnamefont{K.}~\bibnamefont{Sumiyoshi}},
  \bibinfo{journal}{Nucl. Phys.} \textbf{\bibinfo{volume}{A637}},
  \bibinfo{pages}{435} (\bibinfo{year}{1998}), \eprint{nucl-th/9805035}.

\bibitem[{\citenamefont{Fraga et~al.}(2001)\citenamefont{Fraga, Pisarski, and
  Schaffner-Bielich}}]{Fraga:2001id}
\bibinfo{author}{\bibfnamefont{E.~S.} \bibnamefont{Fraga}},
  \bibinfo{author}{\bibfnamefont{R.~D.} \bibnamefont{Pisarski}},
  \bibnamefont{and}
  \bibinfo{author}{\bibfnamefont{J.}~\bibnamefont{Schaffner-Bielich}},
  \bibinfo{journal}{Phys. Rev.} \textbf{\bibinfo{volume}{D63}},
  \bibinfo{pages}{121702} (\bibinfo{year}{2001}), \eprint{hep-ph/0101143}.

\bibitem[{\citenamefont{Alford et~al.}(2005)\citenamefont{Alford, Braby, Paris,
  and Reddy}}]{Alford:2004pf}
\bibinfo{author}{\bibfnamefont{M.}~\bibnamefont{Alford}},
  \bibinfo{author}{\bibfnamefont{M.}~\bibnamefont{Braby}},
  \bibinfo{author}{\bibfnamefont{M.~W.} \bibnamefont{Paris}}, \bibnamefont{and}
  \bibinfo{author}{\bibfnamefont{S.}~\bibnamefont{Reddy}},
  \bibinfo{journal}{Astrophys. J.} \textbf{\bibinfo{volume}{629}},
  \bibinfo{pages}{969} (\bibinfo{year}{2005}), \eprint{nucl-th/0411016}.

\bibitem[{\citenamefont{Demorest et~al.}(2010)\citenamefont{Demorest, Pennucci,
  Ransom, Roberts, and Hessels}}]{Demorest:2010bx}
\bibinfo{author}{\bibfnamefont{P.}~\bibnamefont{Demorest}},
  \bibinfo{author}{\bibfnamefont{T.}~\bibnamefont{Pennucci}},
  \bibinfo{author}{\bibfnamefont{S.}~\bibnamefont{Ransom}},
  \bibinfo{author}{\bibfnamefont{M.}~\bibnamefont{Roberts}}, \bibnamefont{and}
  \bibinfo{author}{\bibfnamefont{J.}~\bibnamefont{Hessels}},
  \bibinfo{journal}{Nature} \textbf{\bibinfo{volume}{467}},
  \bibinfo{pages}{1081} (\bibinfo{year}{2010}), \eprint{1010.5788}.

\bibitem[{\citenamefont{Drago and Tambini}(1999)}]{Drago:1997tn}
\bibinfo{author}{\bibfnamefont{A.}~\bibnamefont{Drago}} \bibnamefont{and}
  \bibinfo{author}{\bibfnamefont{U.}~\bibnamefont{Tambini}},
  \bibinfo{journal}{J. Phys.} \textbf{\bibinfo{volume}{G25}},
  \bibinfo{pages}{971} (\bibinfo{year}{1999}), \eprint{astro-ph/9703138}.

\bibitem[{\citenamefont{Hempel et~al.}(2009)\citenamefont{Hempel, Pagliara, and
  Schaffner-Bielich}}]{Hempel:2009vp}
\bibinfo{author}{\bibfnamefont{M.}~\bibnamefont{Hempel}},
  \bibinfo{author}{\bibfnamefont{G.}~\bibnamefont{Pagliara}}, \bibnamefont{and}
  \bibinfo{author}{\bibfnamefont{J.}~\bibnamefont{Schaffner-Bielich}},
  \bibinfo{journal}{Phys. Rev.} \textbf{\bibinfo{volume}{D80}},
  \bibinfo{pages}{125014} (\bibinfo{year}{2009}), \eprint{0907.2680}.

\bibitem[{\citenamefont{Pagliara et~al.}(2010)\citenamefont{Pagliara, Hempel,
  and Schaffner-Bielich}}]{Pagliara:2010qm}
\bibinfo{author}{\bibfnamefont{G.}~\bibnamefont{Pagliara}},
  \bibinfo{author}{\bibfnamefont{M.}~\bibnamefont{Hempel}}, \bibnamefont{and}
  \bibinfo{author}{\bibfnamefont{J.}~\bibnamefont{Schaffner-Bielich}},
  \bibinfo{journal}{J. Phys.} \textbf{\bibinfo{volume}{G37}},
  \bibinfo{pages}{094065} (\bibinfo{year}{2010}), \eprint{1001.2191}.

\bibitem[{\citenamefont{Pagliara et~al.}(2009)\citenamefont{Pagliara, Hempel,
  and Schaffner-Bielich}}]{Pagliara:2009dg}
\bibinfo{author}{\bibfnamefont{G.}~\bibnamefont{Pagliara}},
  \bibinfo{author}{\bibfnamefont{M.}~\bibnamefont{Hempel}}, \bibnamefont{and}
  \bibinfo{author}{\bibfnamefont{J.}~\bibnamefont{Schaffner-Bielich}},
  \bibinfo{journal}{Phys. Rev. Lett.} \textbf{\bibinfo{volume}{103}},
  \bibinfo{pages}{171102} (\bibinfo{year}{2009}), \eprint{0907.3075}.

\bibitem[{\citenamefont{Prakash et~al.}(1997)}]{Prakash:1996xs}
\bibinfo{author}{\bibfnamefont{M.}~\bibnamefont{Prakash}} \bibnamefont{et~al.},
  \bibinfo{journal}{Phys. Rept.} \textbf{\bibinfo{volume}{280}},
  \bibinfo{pages}{1} (\bibinfo{year}{1997}), \eprint{nucl-th/9603042}.

\bibitem[{\citenamefont{Di~Toro et~al.}(2006)\citenamefont{Di~Toro, Drago,
  Gaitanos, Greco, and Lavagno}}]{DiToro:2006pq}
\bibinfo{author}{\bibfnamefont{M.}~\bibnamefont{Di~Toro}},
  \bibinfo{author}{\bibfnamefont{A.}~\bibnamefont{Drago}},
  \bibinfo{author}{\bibfnamefont{T.}~\bibnamefont{Gaitanos}},
  \bibinfo{author}{\bibfnamefont{V.}~\bibnamefont{Greco}}, \bibnamefont{and}
  \bibinfo{author}{\bibfnamefont{A.}~\bibnamefont{Lavagno}},
  \bibinfo{journal}{Nucl. Phys.} \textbf{\bibinfo{volume}{A775}},
  \bibinfo{pages}{102} (\bibinfo{year}{2006}), \eprint{nucl-th/0602052}.

\bibitem[{\citenamefont{Di~Toro et~al.}(2009)}]{DiToro:2009ig}
\bibinfo{author}{\bibfnamefont{M.}~\bibnamefont{Di~Toro}} \bibnamefont{et~al.}
  (\bibinfo{year}{2009}), \eprint{0909.3247}.

\bibitem[{\citenamefont{Neumann et~al.}(2003)\citenamefont{Neumann, Buballa,
  and Oertel}}]{Neumann:2002jm}
\bibinfo{author}{\bibfnamefont{F.}~\bibnamefont{Neumann}},
  \bibinfo{author}{\bibfnamefont{M.}~\bibnamefont{Buballa}}, \bibnamefont{and}
  \bibinfo{author}{\bibfnamefont{M.}~\bibnamefont{Oertel}},
  \bibinfo{journal}{Nucl. Phys.} \textbf{\bibinfo{volume}{A714}},
  \bibinfo{pages}{481} (\bibinfo{year}{2003}), \eprint{hep-ph/0210078}.

\bibitem[{\citenamefont{Burrows et~al.}(1981)\citenamefont{Burrows, Mazurek,
  and Lattimer}}]{Burrows:1981zz}
\bibinfo{author}{\bibfnamefont{A.}~\bibnamefont{Burrows}},
  \bibinfo{author}{\bibfnamefont{T.~J.} \bibnamefont{Mazurek}},
  \bibnamefont{and} \bibinfo{author}{\bibfnamefont{J.~M.}
  \bibnamefont{Lattimer}}, \bibinfo{journal}{Astrophys. J.}
  \textbf{\bibinfo{volume}{251}}, \bibinfo{pages}{325} (\bibinfo{year}{1981}).

\bibitem[{\citenamefont{Burrows and Lattimer}(1986)}]{Burrows:1986me}
\bibinfo{author}{\bibfnamefont{A.}~\bibnamefont{Burrows}} \bibnamefont{and}
  \bibinfo{author}{\bibfnamefont{J.~M.} \bibnamefont{Lattimer}},
  \bibinfo{journal}{Astrophys. J.} \textbf{\bibinfo{volume}{307}},
  \bibinfo{pages}{178} (\bibinfo{year}{1986}).

\bibitem[{\citenamefont{Keil and Janka}(1995)}]{Keil:1995hw}
\bibinfo{author}{\bibfnamefont{W.}~\bibnamefont{Keil}} \bibnamefont{and}
  \bibinfo{author}{\bibfnamefont{H.~T.} \bibnamefont{Janka}},
  \bibinfo{journal}{Astron. Astrophys.} \textbf{\bibinfo{volume}{296}},
  \bibinfo{pages}{145} (\bibinfo{year}{1995}).

\bibitem[{\citenamefont{Pons et~al.}(1999)\citenamefont{Pons, Reddy, Prakash,
  Lattimer, and Miralles}}]{Pons:1998mm}
\bibinfo{author}{\bibfnamefont{J.~A.} \bibnamefont{Pons}},
  \bibinfo{author}{\bibfnamefont{S.}~\bibnamefont{Reddy}},
  \bibinfo{author}{\bibfnamefont{M.}~\bibnamefont{Prakash}},
  \bibinfo{author}{\bibfnamefont{J.~M.} \bibnamefont{Lattimer}},
  \bibnamefont{and} \bibinfo{author}{\bibfnamefont{J.~A.}
  \bibnamefont{Miralles}}, \bibinfo{journal}{Astrophys. J.}
  \textbf{\bibinfo{volume}{513}}, \bibinfo{pages}{780} (\bibinfo{year}{1999}),
  \eprint{astro-ph/9807040}.

\bibitem[{\citenamefont{Baumgarte et~al.}(1996)\citenamefont{Baumgarte,
  Teukolsky, Shapiro, Janka, and Keil}}]{Baumgarte:1996iu}
\bibinfo{author}{\bibfnamefont{T.~W.} \bibnamefont{Baumgarte}},
  \bibinfo{author}{\bibfnamefont{S.~A.} \bibnamefont{Teukolsky}},
  \bibinfo{author}{\bibfnamefont{S.~L.} \bibnamefont{Shapiro}},
  \bibinfo{author}{\bibfnamefont{H.~T.} \bibnamefont{Janka}}, \bibnamefont{and}
  \bibinfo{author}{\bibfnamefont{W.}~\bibnamefont{Keil}},
  \bibinfo{journal}{Astrophys. J.} \textbf{\bibinfo{volume}{468}},
  \bibinfo{pages}{823} (\bibinfo{year}{1996}).

\bibitem[{\citenamefont{Fischer et~al.}(2009)\citenamefont{Fischer, Whitehouse,
  Mezzacappa, Thielemann, and Liebendorfer}}]{Fischer:2009af}
\bibinfo{author}{\bibfnamefont{T.}~\bibnamefont{Fischer}},
  \bibinfo{author}{\bibfnamefont{S.~C.} \bibnamefont{Whitehouse}},
  \bibinfo{author}{\bibfnamefont{A.}~\bibnamefont{Mezzacappa}},
  \bibinfo{author}{\bibfnamefont{F.~K.} \bibnamefont{Thielemann}},
  \bibnamefont{and}
  \bibinfo{author}{\bibfnamefont{M.}~\bibnamefont{Liebendorfer}},
  \bibinfo{journal}{Astron.Astrophys.} \textbf{\bibinfo{volume}{517}},
  \bibinfo{pages}{A80} (\bibinfo{year}{2009}), \eprint{0908.1871}.

\bibitem[{\citenamefont{Hudepohl et~al.}(2010)\citenamefont{Hudepohl, Muller,
  Janka, Marek, and Raffelt}}]{Huedepohl:2009wh}
\bibinfo{author}{\bibfnamefont{L.}~\bibnamefont{Hudepohl}},
  \bibinfo{author}{\bibfnamefont{B.}~\bibnamefont{Muller}},
  \bibinfo{author}{\bibfnamefont{H.~T.} \bibnamefont{Janka}},
  \bibinfo{author}{\bibfnamefont{A.}~\bibnamefont{Marek}}, \bibnamefont{and}
  \bibinfo{author}{\bibfnamefont{G.~G.} \bibnamefont{Raffelt}},
  \bibinfo{journal}{Phys. Rev. Lett.} \textbf{\bibinfo{volume}{104}},
  \bibinfo{pages}{251101} (\bibinfo{year}{2010}), \eprint{0912.0260}.

\bibitem[{\citenamefont{Pons et~al.}(2000)\citenamefont{Pons, Reddy, Ellis,
  Prakash, and Lattimer}}]{Pons:2000iy}
\bibinfo{author}{\bibfnamefont{J.~A.} \bibnamefont{Pons}},
  \bibinfo{author}{\bibfnamefont{S.}~\bibnamefont{Reddy}},
  \bibinfo{author}{\bibfnamefont{P.~J.} \bibnamefont{Ellis}},
  \bibinfo{author}{\bibfnamefont{M.}~\bibnamefont{Prakash}}, \bibnamefont{and}
  \bibinfo{author}{\bibfnamefont{J.~M.} \bibnamefont{Lattimer}},
  \bibinfo{journal}{Phys. Rev.} \textbf{\bibinfo{volume}{C62}},
  \bibinfo{pages}{035803} (\bibinfo{year}{2000}), \eprint{nucl-th/0003008}.

\bibitem[{\citenamefont{Pons et~al.}(2001)\citenamefont{Pons, Steiner, Prakash,
  and Lattimer}}]{Pons:2001ar}
\bibinfo{author}{\bibfnamefont{J.~A.} \bibnamefont{Pons}},
  \bibinfo{author}{\bibfnamefont{A.~W.} \bibnamefont{Steiner}},
  \bibinfo{author}{\bibfnamefont{M.}~\bibnamefont{Prakash}}, \bibnamefont{and}
  \bibinfo{author}{\bibfnamefont{J.~M.} \bibnamefont{Lattimer}},
  \bibinfo{journal}{Phys. Rev. Lett.} \textbf{\bibinfo{volume}{86}},
  \bibinfo{pages}{5223} (\bibinfo{year}{2001}), \eprint{astro-ph/0102015}.

\bibitem[{\citenamefont{Steiner et~al.}(2001)\citenamefont{Steiner, Prakash,
  and Lattimer}}]{Steiner:2001rp}
\bibinfo{author}{\bibfnamefont{A.~W.} \bibnamefont{Steiner}},
  \bibinfo{author}{\bibfnamefont{M.}~\bibnamefont{Prakash}}, \bibnamefont{and}
  \bibinfo{author}{\bibfnamefont{J.~M.} \bibnamefont{Lattimer}},
  \bibinfo{journal}{Phys. Lett.} \textbf{\bibinfo{volume}{B509}},
  \bibinfo{pages}{10} (\bibinfo{year}{2001}), \eprint{astro-ph/0101566}.

\bibitem[{\citenamefont{Sagert et~al.}(2009)}]{Sagert:2008ka}
\bibinfo{author}{\bibfnamefont{I.}~\bibnamefont{Sagert}} \bibnamefont{et~al.},
  \bibinfo{journal}{Phys. Rev. Lett.} \textbf{\bibinfo{volume}{102}},
  \bibinfo{pages}{081101} (\bibinfo{year}{2009}), \eprint{0809.4225}.

\bibitem[{\citenamefont{Dasgupta et~al.}(2010)}]{Dasgupta:2009yj}
\bibinfo{author}{\bibfnamefont{B.}~\bibnamefont{Dasgupta}}
  \bibnamefont{et~al.}, \bibinfo{journal}{Phys. Rev.}
  \textbf{\bibinfo{volume}{D81}}, \bibinfo{pages}{103005}
  (\bibinfo{year}{2010}), \eprint{0912.2568}.

\bibitem[{\citenamefont{Fischer et~al.}(2010)}]{Fischer:2010wp}
\bibinfo{author}{\bibfnamefont{T.}~\bibnamefont{Fischer}} \bibnamefont{et~al.}
  (\bibinfo{year}{2010}), \eprint{1011.3409}.

\bibitem[{\citenamefont{Mintz et~al.}(2010)\citenamefont{Mintz, Fraga,
  Pagliara, and Schaffner-Bielich}}]{Mintz:2009ay}
\bibinfo{author}{\bibfnamefont{B.~W.} \bibnamefont{Mintz}},
  \bibinfo{author}{\bibfnamefont{E.~S.} \bibnamefont{Fraga}},
  \bibinfo{author}{\bibfnamefont{G.}~\bibnamefont{Pagliara}}, \bibnamefont{and}
  \bibinfo{author}{\bibfnamefont{J.}~\bibnamefont{Schaffner-Bielich}},
  \bibinfo{journal}{Phys. Rev.} \textbf{\bibinfo{volume}{D81}},
  \bibinfo{pages}{123012} (\bibinfo{year}{2010}), \eprint{0910.3927}.

\bibitem[{\citenamefont{Metzger}(2010)}]{Metzger:2010tn}
\bibinfo{author}{\bibfnamefont{B.~D.} \bibnamefont{Metzger}}
  (\bibinfo{year}{2010}), \eprint{1001.5046}.

\bibitem[{\citenamefont{Dall'Osso et~al.}(2010)}]{Dall'Osso:2010ah}
\bibinfo{author}{\bibfnamefont{S.}~\bibnamefont{Dall'Osso}}
  \bibnamefont{et~al.} (\bibinfo{year}{2010}), \eprint{1004.2788}.

\bibitem[{\citenamefont{Bernardini et~al.}(2010)}]{Bernardini:2010zb}
\bibinfo{author}{\bibfnamefont{M.~G.} \bibnamefont{Bernardini}}
  \bibnamefont{et~al.} (\bibinfo{year}{2010}), \eprint{1004.3831}.

\end{thebibliography}

\end{document}